\documentstyle[11pt,aaspp4,psfig]{article}  

\newcommand {\ha} {H$\alpha$\,\,}
\newcommand {\kms} {\,km\,s$^{-1}$\,}

\begin{document}

\title{ 
A massive object in the core of NGC 5055?
}

\author{\sc 
S\'ebastien Blais--Ouellette
}
\affil{
D\'epartement de physique and Observatoire du mont M\'egantic,\\
Universit\'e de Montr\'eal, C.P. 6128, Succ. centre ville,\\
Montr\'eal, Qu\'ebec, Canada. H3C 3J7\\
Electronic-mail: {\it blaisous@astro.umontreal.ca\/}
}



\bigskip
\centerline{\today}

\bigskip

\begin{abstract}
In a global kinematical study of NGC 5055 using high resolution
Fabry-Perot, intriguing spectral line profiles have been observed in the center of
the galaxy. These profiles seem to indicate a rapidly rotating disk with
a radius near
365 pc and tilted 50 $\deg$ with respect to the major axis of 
the galaxy.  In the hypothesis of a massive dark object, a naive
keplerian estimate gives a mass around $10^{7.2}$ to $10^{7.5}$ M$_\odot$.
The limited spectral domain of
the Fabry-Perot leaves some ambiguity on the exact movement and velocity of
this H$\alpha$ emission. 2-D spectroscopy with a larger spectral range
(eg.: TIGRE, OASIS) is thus required.
\end{abstract}

\section{INTRODUCTION}   
\label{intro}            

It is now well establish that many if not all galaxies hide a massive
object in their central region (Kormendy 95). Presence of such objects
are usally 
deduced from the kinematics and photometry of the core of these galaxies.
These Massive Dark Objects (MDO), tought to be black holes, produce 
normally a high velocity dispersion or a rapid rotation
around... nothing  (van der Marel). 

With its high spectral and spatial resolution, the Fabry-Perot
interformeter is well suited for the kinematical study of extended
objects like spiral galaxies. NGC 5055 (M63) is a bright Sbc galaxy
classified as a LINER
in which we wanted to study the detailed kinematical structure of the
H$\alpha$ emission. In the process, our attention have been caught by
the very central part of the galaxy...


\section{OBSERVATIONS}          
\label{obs}    

The Fabry--Perot observations of the \ha emission line were obtained
in 1998 March at the Canada--France--Hawaii Telescope (CFHT).
The Fabry--Perot
etalon (CFHT1) was installed in the CFHT's Multi--Object Spectrograph (MOS).
A narrow--band filter ($\Delta \lambda$ = 10\,\AA), centered at
$\lambda_0$ = 6574\,\AA\, (nearly at the systemic velocity of NGC
5055, V$_{sys} \approx 504$ \kms), was placed in front of the etalon.
The available field with no vignetting was $\approx$ 8.7\arcmin $\times$
8.7\arcmin, with .34\arcsec\, pix $^{-1}$. The free spectral range of
5.66\,\AA\, (258 \kms) was scanned in 28 channels, giving a resolution
per channel of 0.2\,\AA\, (9.2 \kms). 565 seconds integration was
spent at each channel position. 

After reduction (see Amram 91 for more details), we ended up with a 3-D
data set with x,y and $\lambda$ as 
axis. Velocity maps are then elaborated using the intensity
weighted mean of the H$\alpha$ peak to determined its $\lambda$
position thus the radial velocity for each pixel.

\section{KINEMATICS OF THE CENTER}

Globally, the galaxy rotates smoothly and without noticible assymetry
altough some redder flux seems to miss. This could be due to a possible
blueshift of the passband of the (old) filter. The \ha line is normally
symmetrical and well defined where the flux is sufficient.

\begin{figure}                                    
\centerline{\psfig{figure=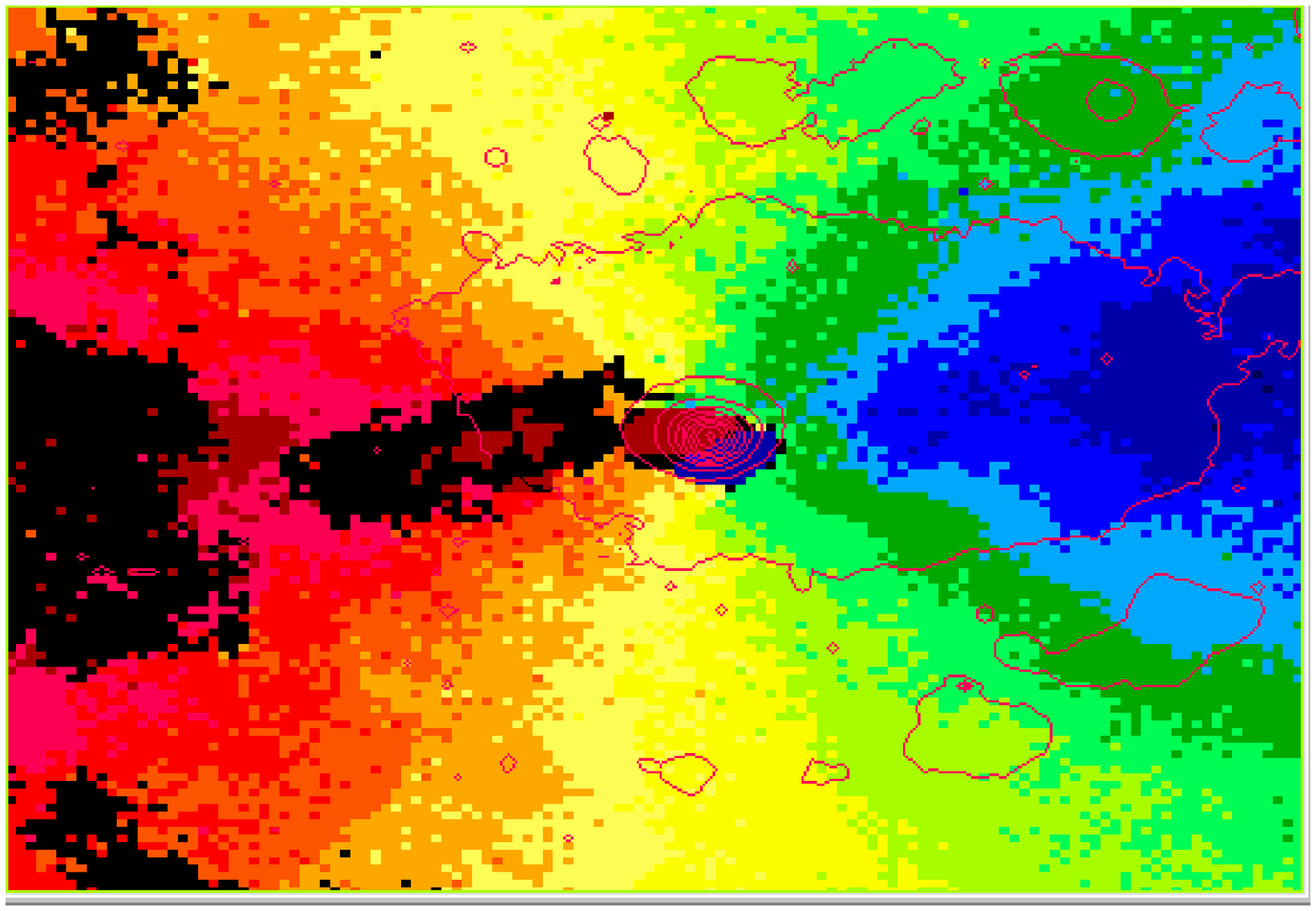,height=3in}} 
\vskip 0.5in
\caption{\label{fig:iso+vf} Isophotes of the integrated flux
superimposed on the velocity field. Line of sight velocity can be 346 or
604 for the south-eastern (blue) blob and 653 or 395 for the
north-western (red).
}
\end{figure}

\begin{figure}                                    
\centerline{\psfig{figure=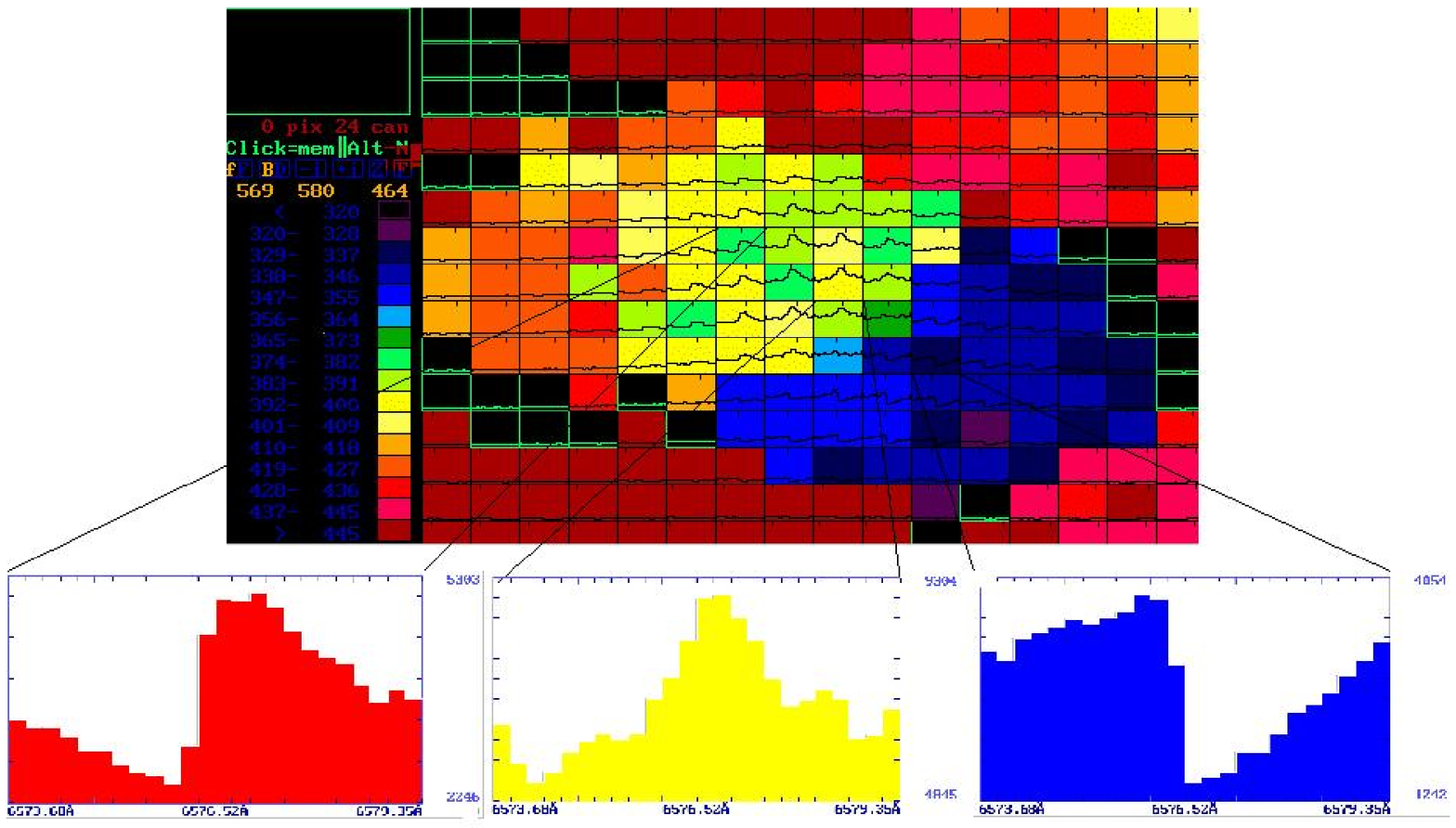,height=3in}} 
\vskip 0.5in
\caption{\label{fig:profiles} Velocity field with typical profile in each region.
}
\end{figure}

When we get to the central 5 arcsecond (\~110 pc), things are changing
radically. In a region where \ha is normally rare, two bright spots
are visible each side of the exact photometric center of the galaxy
(Figure~\ref{fig:iso+vf}). Even more interesting are the
antisymetric appearance of the spectrums of the two
blobs (Figure~\ref{fig:profiles}). When looking at Fabry-Perot
spectrum, one has to keep in mind the intrinsic ambiguity relative to
which interference order we are looking at. If the filter is wide
enough, two or more order can even be superimposed (by slice of 5.67
\AA in this case). This also means that there is a continuity between
the two sides of the spectrum.  

If one look at the profiles in the two spots, one can clearly see a peak
with a long wing on one side and a sharp cut-off on the other. In
between an almost symetrical profile, probably a combination of the
profiles from both side. Because of the ``wraparound'' in the spectrum, it is very
difficult to fix the level of the continuum hence absorption features cannot
be rejected. 

For this central region, velocity have been
calculate as the position of peak in each pixel to avoid being sensitive
to the 
assymetric morphlogy of the peak. To relieve some degeneracy of the
different order of interference, it as been choosen to take one spot
being redshifted from the systemic velocity and the other spot being
blueshifted. Two possibiliies remains. One gives a peak velocity of 653
for the north-western blob and 346 for the south-eastern one. Is also
possible a somewhat conterrotating disk with peak velocity of 395 in the
north-west and 604 in the south-east. Separation between these velocity
are about one arcsecond (37 pc). 

A naive edge-on keplerian model would give for these rotating velocities
between 100 and 150 \kms at 18.5 kpc from the center of rotation, a MDO
mass between $10^7.2 and 7.5 M_{\cdot}$. 

\section{CONCLUSION}

The Fabry-Perot data presented here were optimised for the observation of a large
moderately rotating galaxy. There is thus no surprise if many source of
errors and ambiguities are present when one try to extract valuable
information from a few tens of pixel in a dynamicaly very active
region. 

Obviously, the ambiguity on the real observed wavelenght is very anoying
but managable at the cost of a supplementary hypothesis of a rotation
around the systemic velocity. More damaging is the superposition of many order of
interference since it is forbidding us to fix the real
continuum and ruling out an absorption effect causing the observed
profiles. 

On the other side, the symetrical shapes of the profile clearly indicates that it
is not a systematic error like a drift or a photometric variation. The
high \ha fluxes involved is also a sign that we are in presence of a
quite big amount of energy compatible the presence of a MDO.

Overall, this study shows the necessity of more adapted observations
using integral field spectroscopy where one can trade some field of view
for a larger spectral domain than the Fabry-Perot and a similar resolution. 

\end{document}